\documentclass{article}
\usepackage{PRIMEarxiv}
\usepackage{amsmath}
\usepackage[T1]{fontenc}    
\usepackage{hyperref}       
\usepackage{url}            
\usepackage{booktabs}       
\usepackage{amsfonts}       
\usepackage{nicefrac}       
\usepackage{microtype}      
\usepackage{lipsum}
\usepackage{fancyhdr}       
\usepackage{graphicx}       
\usepackage{xcolor}

\title{Modeling Urban Population Dynamics \\and City-to-City Migration}

\author{
  Rafael Prieto-Curiel\\
  Complexity Science Hub \\ Metternichgasse 8 \\ 1030 Vienna, Austria \\
  \texttt{prieto-curiel@csh.ac.at} \\
   \And
  Carmen Cabrera-Arnau \\
  Department of Geography and Planning \\ University of Liverpool \\ L69 3BX, Liverpool, UK \\
  \texttt{c.cabrera-arnau@liverpool.ac.uk} \\
}

\begin{document}
\maketitle

\section{Abstract}

Migration plays a crucial role in urban growth. Over time, individuals opting to relocate led to vast metropolises like London and Paris during the Industrial Revolution, Shanghai and Karachi during the last decades and thousands of smaller settlements. Here, we analyze the impact that migration has on population redistribution. We use a model of city-to-city migration as a process that occurs within a network, where the nodes represent cities, and the edges correspond to the flux of individuals. We analyze metrics characterizing the urban distribution and show how a slight preference for some destinations might result in the observed distribution of the population.

\section{Introduction}
\begin{figure}
\centering
\includegraphics[width=0.5\textwidth]{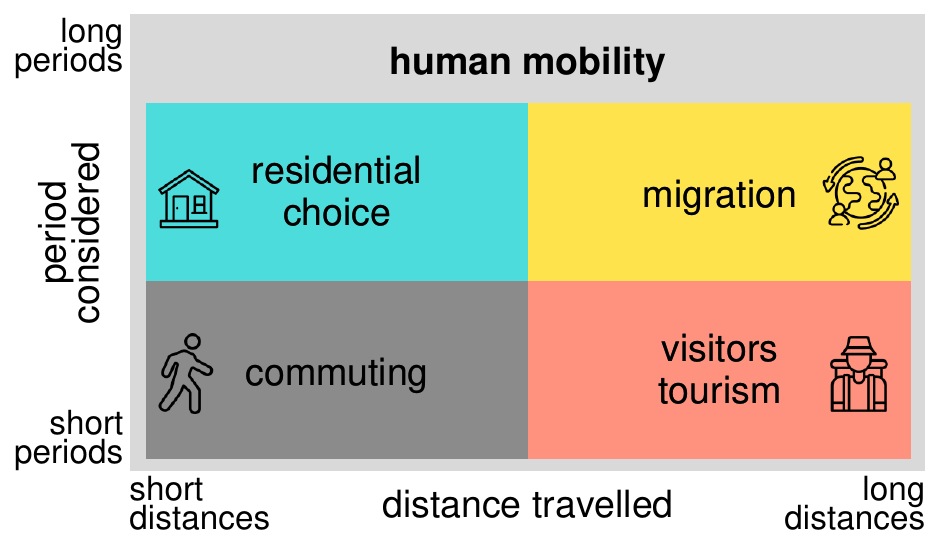}
\caption{Human mobility is a broad concept that includes aspects such as daily commuting, tourism, and migration. Movements that cover short distances and take place for periods shorter than a day are generally associated with daily commutes. Moving home happens over a long period or even permanently and may occur over a wide range of distances. Over short distances, moving home may be a change of residential choice. Over longer distances and long periods, we observe migration.}
\label{flows:MobilityDiagram}
\end{figure}

Human mobility\index{human mobility} takes place at different spatiotemporal scales. Day-to-day mobility, commuting to work or school, leisure, or shopping, takes place at the smallest scale. This type of mobility is characterized by high-frequency and short-distance movements \cite{sirbu2020human, NetworksMigration}. As the frequency lowers and travel distances increase, we may consider visits and tourism. If the time spent at the destination is even more prolonged, the movement may then be a change of place of residence. Migration\index{migration} is observed when the time spent at the destination is prolonged or permanent. Typically, for a movement to be considered migration, a change of residence must occur for at least six months to a year \cite{sirbu2020human, NetworksMigration}. In addition, the new home must be at a long enough distance from the previous one---for example, changing address to a different neighbourhood within the same city is often not regarded as a migratory movement, but changing to a different city is (Fig.~\ref{flows:MobilityDiagram}). Migration, whether international or internal (within the same country), has always been a central feature of humanity, from the first crossing of the Bering Strait to the colonization of America or the rural flight that many countries experienced or are still experiencing during periods of the Industrial Revolution.

Detecting migration patterns formed by people moving between regions within or across countries has been an important research topic for decades. For example, it has been found that the physical distance between an origin and a destination negatively affects the probability of moving, meaning that migrants are more likely to move to destinations relatively closer to their origin \cite{MigrationLaws1885, DistanceOnMigration}. This pattern was already observed more than a century ago by Ernst Georg Ravenstein in the 1880s, described as one of the ``Laws of migration'' \cite{Grigg1977}. More generally, it has been found that the number of visitors to a given location scales with the distance travelled and the frequency of visitation \cite{Schlapfer2021}. As with many other study objects, the statistical significance of observed patterns increases with larger sample sizes \cite{HelbingRybskiEtAl2023}. Traditionally, data on human mobility was only available through administrative data sources such as censuses or surveys. Currently, we also have a range of novel location-tracking data sources that allow us to gather large amounts of information on human mobility and migration at unprecedented spatial and temporal resolution. Examples of data sources that have been used to analyze human mobility and migration include data from social media platforms as well as location tracker devices in mobile phones \cite{Yin17, Yang18, Zhu20, Ponce-Lopez21, Wang22, Rowe23} or smart travel card data \cite{Yang19, Tang20, Nilufer21, Cabrera-Arnau2023}. While these new data sources are very valuable, they also come with challenges, frequently arising from the fact that the data set design was originally not intended for researching human mobility patterns.

Migration can be a highly selective process. For example, economic migrants often migrate for better economic opportunities, such as higher salaries, improved job prospects, or access to resources. Those with the necessary skills, education, and financial means are more likely to migrate for economic reasons. Therefore, economic migrants may have specific characteristics or motivations that set them apart from those remaining in their home location. A similar effect happens with students who move abroad or skilled people moving to a specific industry (for example, Hollywood). As a result, migration shapes the demographics and composition of both the place of origin and the destination, impacting social, economic, and cultural dynamics in both regions. Furthermore, the choice of destination is dependent on the origin of the migrants, whereby international migrants are more likely to move to larger cities. In contrast, rural populations are more likely to move to smaller cities \cite{MigrationLaws1885, ScalingMigrationRPC}.

Population change in a country or a region is driven by migration and natural change\index{natural change}, the latter being the difference between births and deaths. While positive natural change is the primary mechanism driving population growth \index{population growth} in the least developed countries, population change in the most developed countries is either a decline caused by negative natural change or a small growth dominated by incoming migrants. For example, in Djibouti, the total population change in 2021 was 1.37\,\%, and the population change due to natural change was 1.31\,\%. Therefore, natural change accounts for more than 95\,\% of the total growth. In contrast, in Austria, the total population change in 2021 was 0.17\,\%, but the population change due to natural change was -0.04\,\%, so the population would have declined if it was not for the incoming migrants \cite{ourworldindataPopulationGrowth}. While population change is only dominated by migration in some countries, migration always plays the main role in the redistribution of population at the national scale \cite{bell2015internal}. Internal and international migration can alter the population size of cities and, consequently, a country's urban structure\index{urban structure} \cite{bettencourt2020demography, verbavatz2020growth}. Similarly, in urban regions within a country, there is a tendency for population change to be dominated by incoming migrants, whereas natural change tends to be the major driver for population growth in rural regions \cite{MigrationLaws1885}.

Some places ``push'' parts of their population due to unfavourable conditions, whilst other locations attract (or ``pull'') people \cite{lee1966theory}. Then, if this process is sustained consistently over time, the origin and the destination will grow at different speeds. Cities have historically been attractive destinations for migrants, either coming from more rural environments in their relative proximity or from other countries. It is natural that, through the cumulative effect of migration, population growth has historically been faster in cities, leading to more economic activity, infrastructure development, and, overall, increased city attractiveness, resulting in even more migration. This positive feedback cycle can continue for decades or centuries, shaping some cities into massive metropolises with a vast population \cite{Cabrera-Arnau2020}. 

For example, in 1970, the metropolitan area of Delhi had half the population of Kolkata. During the 1960s and 1970s, however, Kolkata was undergoing a time of political instability and financial stagnancy after the partitions of Bengal in 1905 and 1947, worsened by the rise of communal riots and the influx of East Bengali refugees during the Indo-Pakistan war. Combined with the long-term effect of the change of capital city from Kolkata to Delhi in 1911, Kolkata lost its appeal, resulting in high inter-state migration from Eastern and North-Eastern India towards more prosperous regions \cite{internalIndia12}. Today, due to migration, Delhi has twice the population of Kolkata. Similarly, Mexico City has the lowest birth rate in the country, but its population represents a stable 17\,\% of the country's population for more than 50 years, primarily due to migration to the city. In Tanzania, Dar es Salaam has grown fast to 4.5 million people, and currently, it has one-third of the urban population of that country, but it also had one-third of Tanzania's urban population 50 years ago. Dar es Salaam has attracted migration, mainly from the rural parts of Tanzania and other small cities \cite{SecondaryTownsWorldBank}. According to some predictions, Dhaka will have 80 million inhabitants by 2100, and most of this growth will be due to people moving into the city from other regions of Bangladesh \cite{parnell2018}. If Nigeria's population continues to grow and people move to cities at the same rate as now, Lagos could become the world's largest metropolis, home to 85 or 100 million people \cite{100MillionCities}.

Cities such as Paris, Mexico City, or Madrid are large metropolises in France, Mexico, and Spain, respectively, and cities such as Toulouse, Morelia, or Burgos are cities in the same countries of roughly the same age but with much smaller populations. This variation in city sizes\index{city size} is primarily related to migration. Cities may also shrink if they no longer offer good prospects for those moving there. For example, Detroit in the US went from a population of 1.9 million in the 1950s to 0.7 million in 2017. It is mainly due to migration that cities grow to become metropolises (or shrink).

Here, we will provide a general introduction to modelling the evolution of urban population size. Then, we will present a city-to-city migration\index{interurban mobility} model and its implications regarding city size.


\section{A general equation for the evolution of city population size} \label{flows:sec:master-equation}
Generally speaking, population change refers to the change in the number of people in a specified region during a specific period. Population change is usually attributed to two types of change: natural change and migration. Natural change is due to the difference between births and deaths, so if more people are born than dying, we observe positive natural change. Population change due to migration occurs due to the difference between the number of people moving into a location and the number of people leaving that location. So, if more people arrive than leave, there is a positive change due to migration. 

When analyzing population change in the context of urban populations, a set of cities is typically considered as the regions of interest, and time is explored in yearly increments. Any locations outside of the group of cities are deemed to be outside of the system.

Formally speaking, population change is described by the so-called demographic equation\index{demographic equation}. If we consider an urban system formed by $N_c$ cities, where $N_{i}(t)$ is the population of city $i$ at time $t$, then:
\begin{equation}\label{flows:eq:demographic}
    N_i(t+\Delta t)=N_i(t)+\Delta t\left[\upsilon_iN_i(t)+\sum_{j = 1,j\ne i}^{N_c}(J_{ji}(t)-J_{ij}(t))\right]
\, ,
\end{equation}
where $\upsilon_i$ is an effective random growth rate for city $i$, accounting for natural change, i.e.\ population change due to the difference between births and deaths, and for ``out-of-system'' growth, i.e.\ population change due to migratory movements between city $i$ and other places outside the set of $N_c$ cities in the urban system, such as rural regions or cities outside of the region of interest. The terms involving $J_{ij}(t)$ and $J_{ji}(t)$ account for the migratory flows between city $i$ and the rest of the cities in the system. In particular, $J_{ij}(t)$ is the number of people moving from city $i$ to city $j$ between time $t$ and time $t+\Delta t$, and vice-versa for $J_{ji}$. Like in \cite{bettencourt2020demography}, we will work in units where $\Delta t = 1$ for simplicity of notation. 

The model proposed in Eq.~\eqref{flows:eq:demographic} is very general and allows including new, emerging cities over time and excluding others if their size vanishes. In the continuous time limit, Eq.~\eqref{flows:eq:demographic} is equivalent to the diffusion equation with noise, a well-known model with implications far beyond urban systems \cite{Barthelemy2019statistical, verbavatz2020growth} in phenomena such as finance, statistical mechanics, and the Kardar-Parisi-Zhang (KPZ) equation. However, as is often the case when using very general equations, it is challenging to use Eq.~\eqref{flows:eq:demographic} for making predictions \cite{verbavatz2020growth}, as there are no known solutions for a general form of $J_{ij}$ \cite{Barthelemy2019statistical}. Various assumptions have been proposed to simplify the equation, leading to other well-known models for urban growth. For example, when we assume that there is an exact balance of migration flows, i.e.\ $J_{ij} = J_{ji}$, Eq.~\eqref{flows:eq:demographic} becomes equivalent to Gibrat's model \cite{gibrat1931inegalites}, which predicts a log-normal distribution of city population sizes. Similarly, when the mean-field approximation is considered, i.e.\ $J_{ij} = J/N_c$ for all $i,j=1, \dots, N_c$ with $J =\sum_{i,j=1}^{N_c}J_{ij}$, and we assume that $J$ is small enough, then the solution of Eq.~\eqref{flows:eq:demographic} leads to a power law distribution in equilibrium, which is consistent with the empirical Zipf's law \index{Zipf's law}. This suggests that the origin of Zipf's law lies in the interplay between internal random growth and exchanges between different cities. However, according to the mean-field approximation, as interurban mobility\index{interurban mobility} increases, the population sizes of all the cities in the system converge to the same value. Therefore, from a theoretical perspective, it is essential to consider not just the mean-field analysis but also other non-constant coupling $J_{ij}$ alternatives \cite{Barthelemy2019statistical} that are capable of accounting for random fluctuations in the migratory flows migrations.

In the remainder of this chapter, we discuss several models of city-to-city migration that go beyond the mean-field approximation. For simplicity, we will set $\nu_i = 0$ and focus only on urban population growth purely due to intraurban migration flows, so instead of working with Eq.~\eqref{flows:eq:demographic}, we will consider the following equation
\begin{equation}\label{flows:eq:internal}
    N_{i}(t+1) = N_i(t) + \sum_{j = 1, j\ne i}^{N_c}(J_{ji}(t)-J_{ij}(t))
    \, .
\end{equation}

\section{Modeling city-to-city migration}
Network analysis\index{network analysis} offers a simple way to conceptualize migratory movements between cities. Here, we represent the distinct origin and destination locations as network nodes, in our case, cities. The weights of the edges represent the probability that a person moves between the two nodes. This type of network has been used to model migration between California and other states in the US or people moving from and to Germany \cite{MarkovMigration, FiniteMobility, PrietoColombiaMobility, RepeatMigrationMarkov}. 

Using the same notation introduced in Sec.~\ref{flows:sec:master-equation}, we consider a region with $N_c$ cities. Usually, to keep the notation simple, we assign the same index $i={1, 2, \dots, N_c}$ to each node and to the city it represents so that node 1 corresponds to ``city 1'' and node 4 to ``city 4''. We add weighted edges between the pairs of nodes. The weight of an edge between nodes $i$ and $j$ corresponds to the probability $m_{ij} \geq 0$ that a person moves between cities $i$ and $j$. Migration is a process that takes a long time, so in general, the movements between cities are considered over annual intervals as opposed to daily or weekly intervals. If $m_{ij}=0$, we consider that the edge between $i$ and $j$ does not exist. Let $m_{ii}$ be the probability that the person remains in the city $i$ from one year to the next. This way, the sum of the probabilities satisfies $\sum_{j \in n} m_{ij} = 1$, for $i=1,2, \dots N_c$. Here, we assume that the probability of moving is independent of past migrations, that it is the same for all individuals, and that it only depends on the origin and destination cities of a possible migratory movement.

Then, the number of people that move from $i$ to $j$ in a year, which we denote by $J_{ij}(t)$, can be regarded as a time-dependent random variable with an associated probability mass function. Even though the variables corresponding to the cities' populations and the intraurban population flows are time-dependent, we will not make this explicit for ease of notation unless it is necessary for clarity.

Since $J_{ij}$ represents a count, a natural choice of distribution to model this quantity is a Binomial distribution. A Binomial distribution is often used to model the probability for a certain number of successes, given a number of random trials and the probability for success in each individual trial. We can think of the number of people in $i$, $N_i$, as the number of trials, and the probability of an individual migrating from $i$ to $j$, $m_{ij}$, as the probability for success in each trial. Then, the number of people migrating from $i$ to $j$, $J_{ij}$, can be regarded as the number of successes, distributed according to a Binomial distribution of the form
\begin{equation}
\begin{aligned}
   \text{Pr}(J_{ij}; N_i, m_{ij}) = \text{Bin}(J_{ij},N_i, m_{ij}) \\
   = {{N_i}\choose{J_{ij}}}m_{ij}^{J_{ij}}(1-m_{ij})^{N_i-J_{ij}}
   \, .
\end{aligned}
\nonumber
\end{equation}

The expected number of people moving from $i$ to $j$ is $E[J_{ij}] = m_{ij} N_i$, corresponding to the expected value of the Binomial distribution, and the variance is $Var[J_{ij}] = m_{ij}(1-m_{ij}) N_i$. Note that the expected value of $J_{ij}$ depends on the population size of the origin node. Thus, considering the interaction of $i$ only with node $j\neq i$, the expected population of city $i$ at time $t+1$ is $N_i(t+1) = N_i(t) + m_{ji}N_{j}(t) - m_{ij}N_i(t)$. 

If, instead, we consider the interaction of $i$ with all the other nodes in the network (i.e.\ the flows between city $i$ and all the other cities in the urban system), the model can be generalized from a Binomial to a Multinomial distribution. The mathematical properties are pretty similar \cite{Seber2011multinomial}. The joint probability of the flows of people moving from node $i$ to all other nodes in the network, given a population $N_i$ at the origin and a set of probabilities $\mathbf{m}_i = (m_{i1}, ...,m_{iN_c})$ for moving from $i$ to each $j$ in the network, can be modelled as follows
\begin{equation}\label{flows:eq:multinomial}
\begin{split}
    Pr(\mathbf{J}_i;N_i,\mathbf{m}_i) &=\\
    &=\text{Multi}(\mathbf{J}_i,N_i,\mathbf{m}_i) \\
    &=\dfrac{N_i!}{J_{i1}! \dots J_{iN_c}!}m_{i1}^{J_{i1}}\times \dots \times m_{iN_c}^{J_{iN_c}} \, ,
\end{split}
\end{equation}
where $\mathbf{J}_i$ represents the set of flows from $i$ to every other node including itself, $(J_{i1}, \dots, J_{iN_c})$.


Underlying Eq.~\eqref{flows:eq:internal}, there is a discrete-time Markov chain\index{Markov chain} that models the states of a random variable over time. An important feature of Markov chains is that the current state of the variable depends only on the state at the previous time step \cite{Blanchard2011rwgraph}. In our case, the random variable is categorical and indicates the city where an individual is located. The possible states of this variable are the cities in the network. Hence, the Markov chain describes the sequence of cities in the network that the individual would visit at each time step. The probability of visiting city $i$ at time $t$ only depends on the city where they were located at the previous time step. Generally, a Markov chain can be characterized by a stochastic matrix, which contains the probabilities that the random variable transitions from one state to another in each time step \cite{Blanchard2011rwgraph}. Here, this matrix $M$ is given by the probabilities to migrate between from city $i$ to any other city $j$
\begin{equation}
\label{flows:matrix}\nonumber
    M = \begin{pmatrix}
m_{11} & m_{12} & \dots & m_{1N_c}\\
m_{21} & m_{22} & \dots & m_{2N_c}\\
\vdots & \vdots & \ddots & \vdots \\
m_{N_c1} & m_{N_c2} & ... & m_{N_cN_c}
\end{pmatrix}
\, .
\end{equation}
Note that, in general, these transition probabilities are not necessarily independent of time. If they are, we say that the Markov chain is time-homogeneous. 
The total population of the system, i.e.\ the sum of the population of all the $N_c$ cities, is, on average, conserved over time. This is ensured by the fact that the elements in each row of $M$ add up to 1. We can prove this by showing that, at any time $t$, the equality $E[\sum_{i}N_i(t+1)] = \sum_{i}N_i(t)$ is satisfied
\begingroup
\allowdisplaybreaks
\begin{align}\label{flows:eq:pop_conservation}
    E&[\sum_{i=1}^{N_c}N_i(t+1)] = \sum_{i=1}^{N_c}E[N_i(t+1)] = \\
    &= \sum_{i=1}^{N_c}N_i(t) + \sum_{i=1}^{N_c}\sum_{\substack{j=1\\ j\neq i}}^{N_c}m_{ij}N_j(t) - \sum_{i=1}^{N_c}\sum_{\substack{j=1\\ j\neq i}}^{N_c}m_{ij}N_i(t)\nonumber = \sum_{i=1}^{N_c}N_i(t) \, . \nonumber
\end{align}
\endgroup 
where the two last terms in the second last step add up to 0, as it can be derived from the fact that $\sum_{i}\sum_{j, j\neq i}m_{ij}N_j(t) = \sum_{i}\sum_{j}m_{ij}N_j(t) - \sum_{i}m_{ii}N_i(t) = \sum_{i}\sum_{j}m_{ij}N_j(t) = \sum_{j}N_j(t)\big[\sum_{i}m_{ji}\big] = \sum_{j}N_j(t)$ and equivalently for the subtracting term, $\sum_{i}\sum_{j, j\neq i}m_{ij}N_i(t)$.

It is possible to run simulations using an iterative algorithm, where each iteration represents a time step of the stochastic process. We set the initial time of this process to $t=0$. At $t=0$, the population size of all cities is known, and it is given by $\mathbf{N}(0) = (N_1(0), \dots, N_{N_c}(0))$, and the matrix of transition probabilities $M$ is also known. Let's assume the Markov process is time-homogeneous, so the matrix $M$ remains constant over time. The algorithm involves the following steps, starting from $t=0$:
\begin{enumerate}
    \item For each node $i$, generate the number of people moving to every other node in the network, $J_{ij}$. This will be a set of random numbers drawn from a Multinomial distribution specified in Eq.~\eqref{flows:eq:multinomial}, parametrised by $N_i$ and $m_i$.
    \item For each node $i$, update the corresponding city population to generate $N_i(t+1)$. This is done by subtracting from $N_i(t)$ the total number of people that move to other nodes, i.e.\ $\sum_{j=1, j\neq i}^{N_c}J_{ij}$ and by adding to the resulting quantity the total number of people that move from other nodes, i.e.\ $\sum_{j=1, j\neq i}^{N_c}J_{ji}$.
    \item Update $t$ to $t+1$ and repeat from step 1.
\end{enumerate}

To obtain confidence intervals for the city population sizes at time $t$, we can run the simulations a sufficiently large number of times up to time $t$. Although there is no specific requirement for the number of iterations, at least 100 simulations are frequently used.

\section{Characterizing the disparity in the distribution of city population sizes at time $t$}

The size distribution of many events in nature is characterized by a few huge observations and many smaller ones. For example, thousands of low-magnitude earthquakes have been recorded over a century, with very few earthquakes of magnitude eight or above. Similarly, there are only one or two massive cities in a country or region, whereas there are many smaller ones \cite{Maitre1682, Auerbach31}. Heavy-tailed probability distributions of event sizes, such as a power law or a lognormal distribution, are often used to account for this pattern. Zipf's law \cite{Zipf49} is an empirical law that states that city population size follows a power law distribution with exponent -1, and it has repeatedly been reported to hold in the case of the distribution of urban population sizes (e.g.\ \cite{Ioannides03Zipfs}). In the context of the distribution of wealth, Pareto's distribution is frequently used to describe that a small number of individuals concentrate a large proportion of the total wealth.

The Gini index\index{Gini index} is one of the most widely used measures to quantify the disparity in the distribution of wealth. It is possible to establish explicit functional relations between the parameters that characterize the probability distribution of the variable of interest and the Gini index \cite{pfahler85RelativeConcentration}. However, this index may also be used to quantify the disparity in the sizes of other variables, such as city population size. Like the Gini index, other metrics summarize the disparity of a distribution. To be able to compare distributions across countries or regions and over time, disparity metrics need to satisfy the following properties \cite{Cowell00}:
\begin{itemize}
    \item Anonymity principle: permutations of ``personal'' labels within a given distribution should not affect the overall level of disparity, e.g.\ if the population of London and Manchester are swapped, the level of disparity for city population size in England should remain the same.
    \item Transfer principle (or Pigou-Dalton principle): disparity is reduced when a fixed amount is transferred from a larger to a smaller event, and the recipient is still smaller than the donor. For example, if people from the capital migrate to a smaller town, the level of disparity would be decreased.
    \item Population principle: the level of disparity should be independent of the size of the data set (e.g.\ number of cities in the sample). This ensures that it is possible to compare small and large regions or countries regardless of their total number of cities.
    \item Relative size principle: only relative values should matter. If the values for city population size increase by a constant proportion for all the cities in the dataset, then the level of disparity should remain the same. For example, if all cities in a country grew in size by the same proportion due to an overall increase in fertility rates, the level of disparity should remain the same.
\end{itemize}

Here, we focus on four metrics of disparity in city population size. These are (i) the Gini index, (ii) the metropolization index, (iii) the 90-10 inter-decile ratio, and (iv) the coefficient of variation. While the Gini index and the coefficient of variation use the whole distribution of city population sizes, the metropolization index and the 90-10 inter-decile ratio only look at specific distribution points. 

Let us start by defining the Gini index. Let $N_i$ be the population size of city $i$, with $i=1, \dots, N_c$, then the Gini index $G$ is given by
\begin{equation}
\label{flows:Gini}\nonumber
    G = \dfrac{\sum_{i=1}^{N_c}\sum_{j=1}^{N_c}|N_i - N_j|}{2\sum_{i=1}^{N_c}\sum_{j=1}^{N_c}N_j} \, .
\end{equation}
The Gini index's range of values spans from 0 to 1, with these extremes corresponding, respectively, to a situation where all the cities have the same population size and to a situation of maximal disparity in population sizes.

The metropolization index $MI$ is defined here as the proportion of the population living in the largest city, i.e.
\begin{equation}
    MI = \dfrac{max (\{N_i\}_{i=1}^{N_c})}{\sum_{i=1}^{N_c} N_i} \, ,
    \nonumber
\end{equation}
with values of $MI$ varying between $1/{N_c}$ if all cities have the same population and a value close to 1 if one city has most of the region's population.

The 90-10 inter-decile ratio $IR$ measures how many times larger the population size at the 90th percentile, denoted by $N_{\%90}$, is compared with the population size at the 10th percentile, denoted by $N_{\%10}$. Hence,
\begin{equation}
    IR = \dfrac{N_{\%90}}{N_{\%10}} \, .
    \nonumber
\end{equation}
This measure takes a minimum value of 1, corresponding to the scenario where all the cities have the same population. It tends to infinity as the disparity in the distribution of city sizes increases.

Finally, the coefficient of variation $CV$ equals the ratio of the standard deviation to the average of the city population size. If only a sample of empirical data is available, the CV can be estimated using the ratio of the sample standard deviation to the sample mean, i.e., 
\begin{equation}
    CV = \dfrac{\sqrt{\dfrac{1}{N_c}\sum_{i=1}^{N_c}(N_i-\bar{N})}}{\bar{N}}\, ,
    \nonumber
\end{equation}
where $\bar{N} = \frac{1}{N_c}\sum_{i=1}^{N_c}N_i$ is the sample mean. The $CV$ measures variability concerning the mean, so it is independent of the values of city population size and, therefore, satisfies the "population principle". While the Gini index is more sensitive to changes in the middle values of the distribution of sizes, the $CV$ places more weight on the top end of the distribution. Therefore, if one of the largest cities in the region under study gains population faster than the others, the change in value in $CV$ will be relatively greater than the Gini index.

\section{Migration between five cities}
\begin{figure}
\centering

\includegraphics[width=0.6\textwidth]{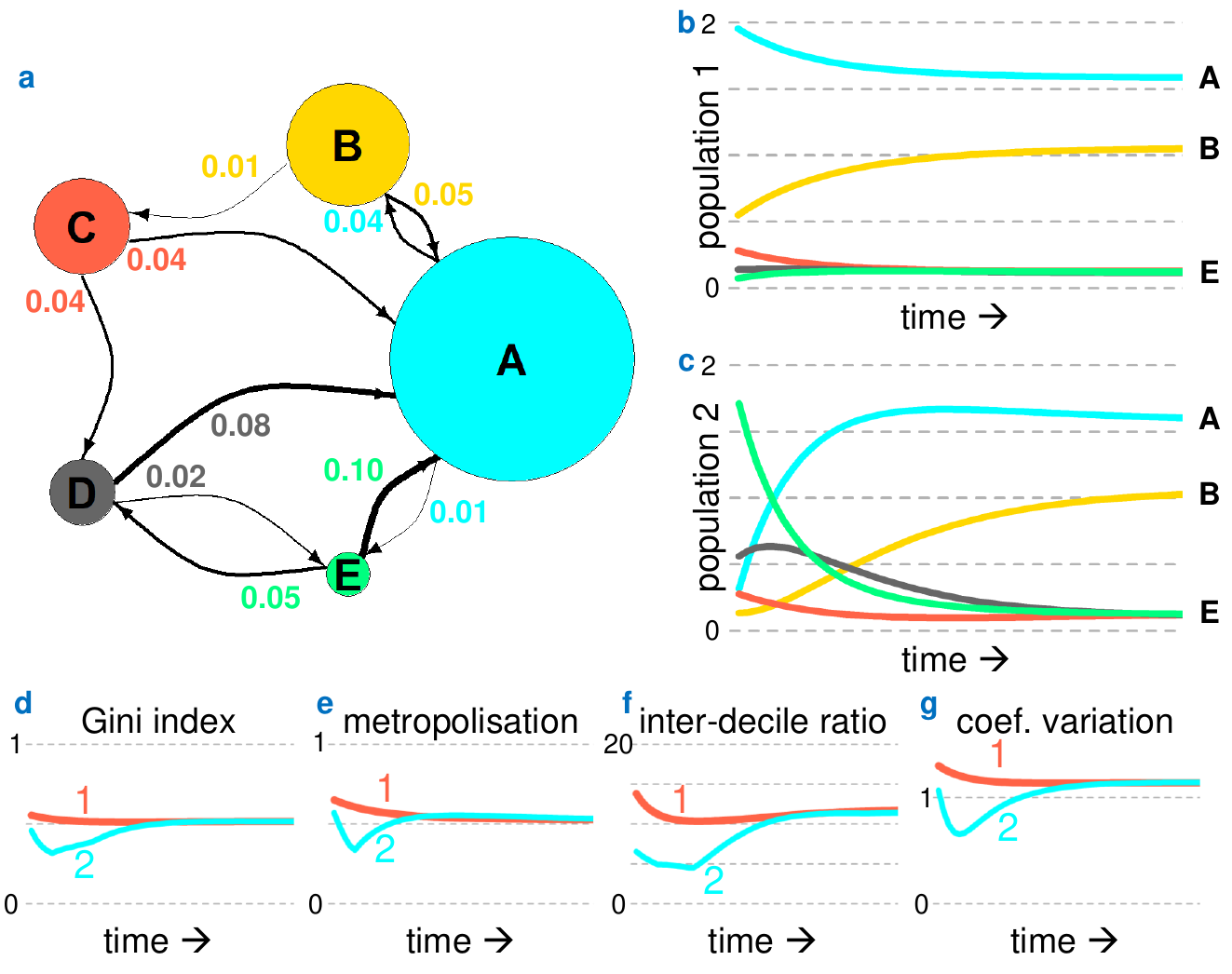}
\caption{(a) Five nodes represent five cities, labelled $A$, $B$, $C$, $D$ and $E$. The edges have been given probabilities according to Tab.~\ref{flows:tableProbs}. The nodes' size on the left diagram corresponds to the long-run distribution of the population. (b,c) Evolution of city size, computed according to Eq.~\eqref{flows:eq:demographic} for different initial distributions of population sizes $P(0)$. After a few time steps, both distributions (b and c) are roughly the same. Evolution of the Gini index (d), the metropolisation index (e), the 90-10 inter-decile ratio (f), and the coefficient of variation (g) for the two initial distributions of populations of sizes.} 
\label{flows:FiveNodesFig}
\end{figure}

Consider a network with five nodes corresponding to a system of five cities (Fig.~\ref{flows:FiveNodesFig}). The matrix $M$ in Tab.~\ref{flows:tableProbs} gives the probability of moving between cities.

\begin{table}[h!]
\caption{Probability of moving between two cities. The origin corresponds to the rows of the table, and the destination corresponds to the columns. Thus, the probability of moving between $A$ and $E$, for example, is $m_{AE}=0.01$.}
\label{flows:tableProbs}
\centering
\begin{tabular}{c ccccc} 
\hline
& $A$&$B$&$C$&$D$&$E$\\
\hline
$A$ & 0.95 & 0.04 & 0    & 0    & 0.01 \\
$B$ & 0.05 & 0.94 & 0.01 & 0    & 0    \\
$C$ & 0.04 & 0    & 0.92 & 0.04 & 0    \\
$D$ & 0.08 & 0    & 0    & 0.90 & 0.02 \\
$E$ & 0.10 & 0    & 0    & 0.05 & 0.85 \\
\hline
\end{tabular}
\end{table}

Migration from city $A$ to city $B$ differs from migration from $B$ to $A$. This is reflected by the matrix $M$ is not symmetric, meaning that $m_{ij}$ is not always the same as $m_{ji}$. Some cities could attract more people than others. Note, however, that the total number of people in our system of cities is, on average, constant over time since we are not considering births, deaths or out-of-system population growth. This is ensured by the fact that the values in each row of matrix $M$ add up to one, as demonstrated in Eq.~\eqref{flows:eq:pop_conservation}. For example, in Tab.~\ref{flows:tableProbs}, the first row gives the probabilities of moving from city $A$ to each other city, and the sum is 1. The first column shows the probabilities of moving to city $A$ from any other city, and these values are higher than for moving from $A$ to any other city. That is, $m_{iA} > m_{Ai}$, with $i=B, C, D, E$, which means that city $A$ pulls more people from cities $B, C, D$ and $E$ than it pushes to those cities.

Taking some initial population $P(0)$, we can analyze the evolution of the size of all cities (Fig.~\ref{flows:FiveNodesFig}, right). With different initial populations (top and bottom of Fig.~\ref{flows:FiveNodesFig}), the curves representing the evolution of city population sizes have different shapes. However, after enough time steps, the distribution of sizes becomes nearly the same. As time passes, the population flows between cities become more dominant than the initial populations in determining the long-term distribution of population sizes. In both cases depicted in Fig.~\ref{flows:FiveNodesFig}, city $A$ becomes the most populous, followed by city $B$. We can attribute this to the fact that, according to the transition probabilities in matrix $M$, people are more attracted to city $A$ and are less likely to move outside that city. Moreover, the values in the diagonal of $M$ are relatively high, corresponding to the probability of a person staying in the same city from one year to the next. This will be the case in general as most people will remain in the same city from one year to the next, and in relative terms, almost no one will move. Many people will never move from their hometown, and those who do will likely remain in their new location for years, or even decades, before moving again \cite{PrietoColombiaMobility}.

The impact of the initial size of the cities becomes less relevant as time passes. Even though there are differences in the distribution of the populations in the first stages, the long-run distribution of the population is the same for both sets of initial city population sizes (Fig.~\ref{flows:FiveNodesFig}). Therefore, the Gini index, the metropolization index, the inter-decile ratio, and the coefficient of variation also converge to similar values. In the example, for the initial city population sizes plotted on top (population 1), the Gini index is relatively high, meaning that a few cities have most of the country's population; the metropolization index is above 0.5, meaning that the largest city has more than half of the population; the inter-decile ratio is above 10, and the coefficient of variation is above 1. The initial city population sizes plotted on the bottom (population 2) are more mixed, so there are some short-term fluctuations in the four metrics under consideration. However, in the long run, the values converge to those obtained with the first set of initial city population sizes.

\section{Modeling internal migration in Madagascar}
\begin{figure}
\centering

\includegraphics[width=0.6\textwidth]{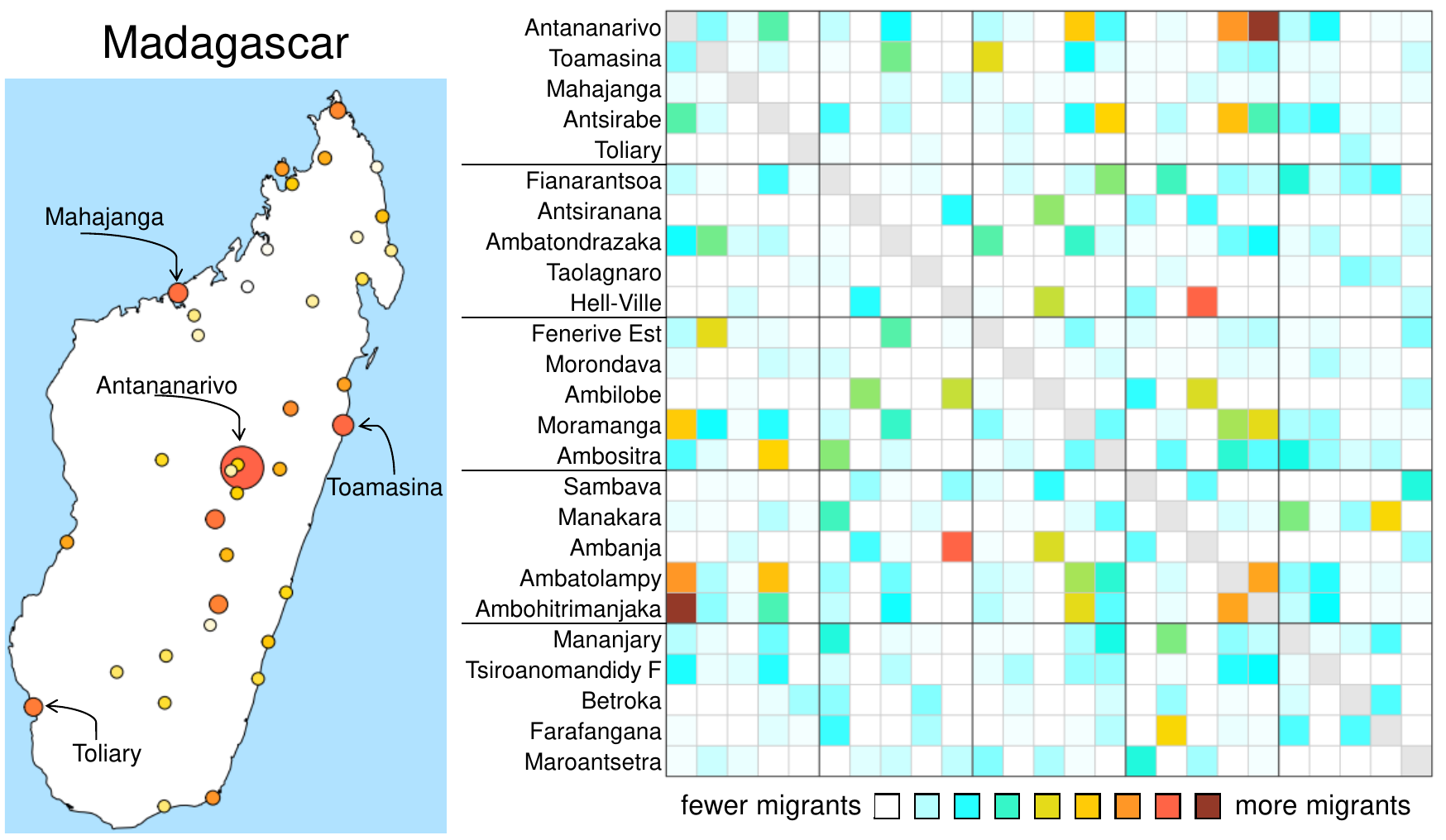}
\caption{Madagascar example. The map of Madagascar and its 25 most populous cities is shown on the left. The disk's size corresponds to each city's population in 2015. On the right, the probability of moving between two cities $m_{ij}$ considering only the distance between city $i$ and $j$, with $\alpha = 0.04$ and $\beta = 1/100$, is illustrated. Columns and rows are sorted according to the population of cities.}
\label{flows:MadagascarMap}
\end{figure}

Here, we construct an example in Madagascar, a country occupying an island in Southeast Africa with roughly the same surface as France. Madagascar has approximately 30 million inhabitants, with its capital being Antananarivo, with 2.5 million people. The Africapolis dataset gives an estimate of the population of urban agglomerations in the continent, and it will be used to measure population and distance between cities \cite{Africapolis}. Madagascar has 68 urban agglomerations, but here we focus on the top 25 most populous cities (Fig.~\ref{flows:MadagascarMap}).

First, let us consider what happens if people move to cities depending only on the distance between their city and the possible destinations. The physical distance between city $i$ and $j$ reduces the probability of moving. One way to capture this process is by considering the distance between the two cities, $D_{ij}$, and an exponential decay probability. That is, we can capture the effect of distance by expressing the probability of moving by
\begin{equation}
    m_{ij} = \alpha \exp(- \beta D_{ij})\, ,
    \nonumber
\end{equation}
where $\alpha>0$ and $\beta>0$ are model parameters. Notice that due to the negative sign inside the exponential, the expression gives a smaller probability of moving for longer distances. For example, with values of $\alpha = 0.04$ and $\beta = 1/100$, we obtain the matrix $M$ in Fig.~\ref{flows:MadagascarMap}. For a given $i$, the distance from $i$ to itself is $D_{ii}=0$, which means that $\alpha=m_{ii}$. Furthermore, according to a Multinomial distribution, the values $m_{ij}$ for $j=1, \dots, N_c$, must add up to 1. Therefore, the value of $\beta$ must be such that for any $i$, $\sum_{j=1}^{N_c}m_{ij} = 1$. 

\begin{figure}
\centering

\includegraphics[width=0.6\textwidth]{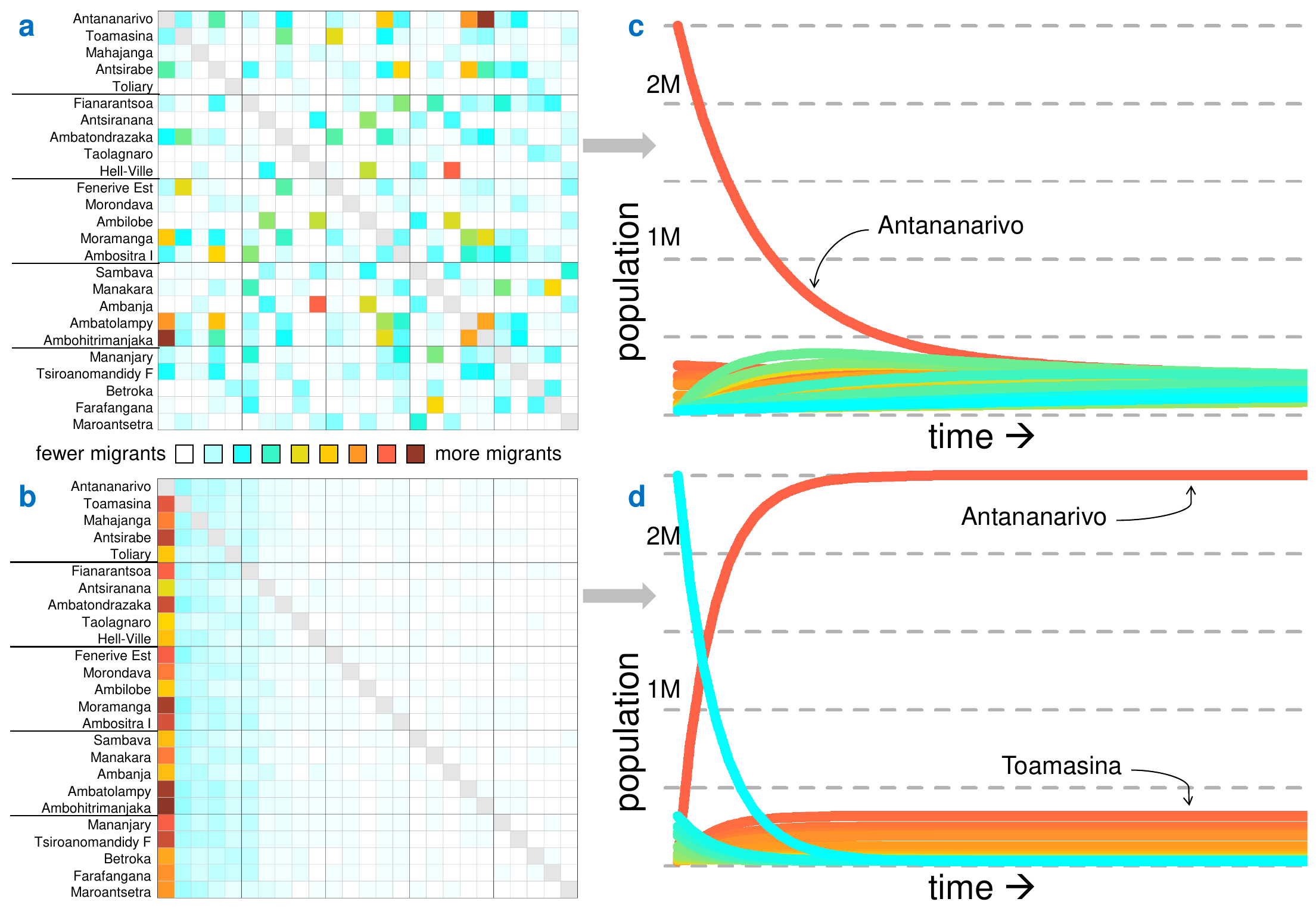}
\caption{Modeled probability of moving between cities (left) and distribution of city size as time evolves (right). The top panel (a) shows the scenario where the distance gives the probability of moving between cities, i.e.\ $m_{ij} = \alpha \exp(- \beta D_{ij})$. The bottom panel (b) shows the scenario where the probability is dependent on destination size too, i.e.\ $m_{ij} = \gamma N_j \exp(- \delta D_{ij})$. In both cases, the country's total population remains constant over time, so $\sum_{i=1}^{N_c}N_i(t) = \sum_{i=1}^{N_c}N_i(t + 1)$ for any $t\geq 0$.} 
\label{flows:MadagascarImpact}
\end{figure}

The case where the probability of moving between cities is modelled using only distance has some interesting properties. First, the matrix $M$ is symmetric (since distances are symmetric). Hence, if city $i$ has more population than $j$, i.e.\ $N_i(t) > N_j(t)$, then the expected number of people moving between cities $i$ and $j$, $E[J_{ij}]$ exceeds the expected number of people moving between city $j$ and $i$, $E[J_{ji}]$ since according to the Multinomial distribution $E[J_{ij}] = m_{ij}N_{i}$, therefore:
\begin{equation}
E[J_{ij}] =  m_{ij} N_i(t) >  m_{ji} N_j(t) = E[J_{ji}]
\, .
\nonumber
\end{equation}
The expected population flow between cities $i$ and $j$ would be the same as the flow between cities $j$ and $i$ only if both cities have the same population. Otherwise, more people will move from large to small cities. In the long run, an equilibrium will be reached when all cities have the same population. This model is illustrated in the top panel of Fig.~\ref{flows:MadagascarImpact}). 

While this model works mathematically, it leads to an equilibrium distribution that fails to capture the disparity in city population sizes in the long run. A more realistic approach (as we will see below) is to model the transition probabilities $m_{ij}$ as a function of the distance between cities and the population of $N_j$ rather than $N_i$, hence placing more emphasis on the characteristics of the possible destination locations that an individual might decide to migrate to.

Considering $m_{ij} = \gamma N_j \exp(- \delta D_{ij})$, with $\gamma >0$ and $\delta > 0$, the long-run distribution of the population is similar to the one observed in Madagascar (Fig.~\ref{flows:MadagascarImpact}), i.e.\ very few large cities and lots of smaller cities. In this case, the initial distribution of the population at time $t=0$ becomes less relevant, and people's preference for particular destinations determines which cities will eventually be the largest ones in the country. Here, we assume that births and deaths are the same, so the only demographic process we observe is migration between cities. Here, if people are slightly more likely to move to a city like Antananarivo, it will eventually become the largest city in the country.

\begin{figure}
\centering

\includegraphics[width=0.6\textwidth]{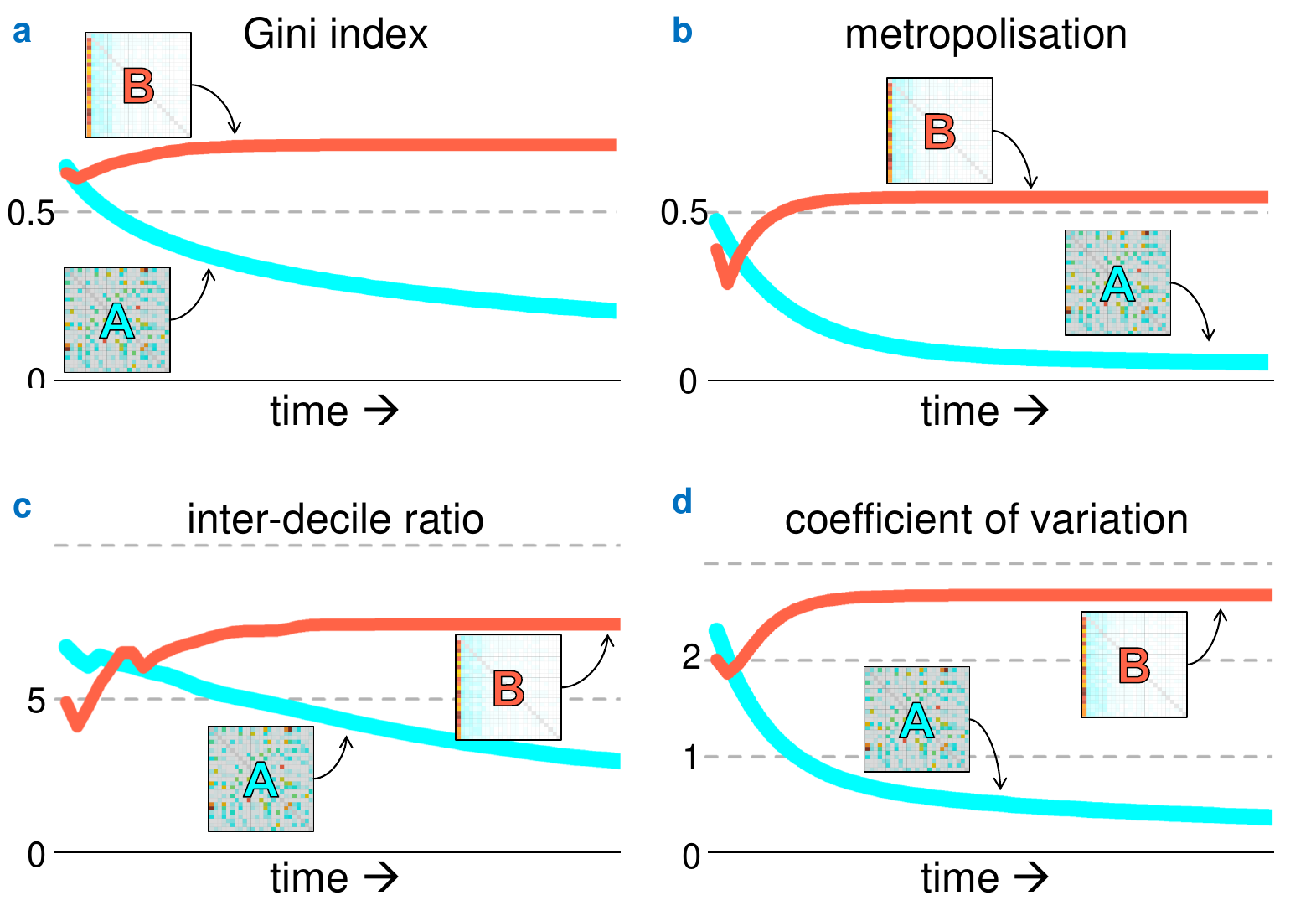}
\caption{Temporal evolution of the Gini index (a), the metropolization index (b), the inter-decile ratio (c), and the coefficient of variation (d), according to two migration models. The first is based on proximity between cities A or on proximity and population size of the destination B. The matrices correspond to the tables in Fig.~\ref{flows:MadagascarImpact} on top A and on bottom B.} 
\label{flows:MadagascarMetrics}
\end{figure}

The metrics for the population distribution show substantial differences (Fig.~\ref{flows:MadagascarMetrics}). Suppose large cities attract more population (second model). In that case, the Gini index\index{Gini index} and the coefficient of variation are higher, showing the existence of more disparity in the distribution of city population sizes, with just a few large cities hosting most of the population. Also, the metropolization index is higher, indicating that this concentration is observed mainly in a single city. Finally, the coefficient of variation is also high, meaning that the size of cities is very unequal. On the contrary, in the model based on proximity only, the Gini coefficient drops as time passes, meaning that cities evolve to a more similar size. The metropolization index also falls, meaning that no city absorbs most of the population in that system, and the 90-10 inter-decile ratio also drops, showing that the top and bottom cities have a similar size. Finally, a slight variation coefficient means the size distribution is less skewed.

\section{Conclusion}
Human mobility is a broad topic, encompassing daily commuters, residential changes, tourism, visits, and migration. Migration is observed only when the person has moved a sufficiently large distance and for an extended period. Migration profoundly impacts society, including political expressions, use of services, demographic processes, and others. Here, we have discussed only observed migration and its implication on altering city size.

There are many ways to model migration. Our proposed models, where we consider the possible locations as the nodes in a network of cities and the probability of moving between cities as the weighted edges, are just some among many. Our approach relies on a Markov process, which describes the changes over time of an individual's location in the network according to the probability given by the edges' weights. The approach also relies on the assumptions that people move depending only on features of their current location or possible destinations and that their movement does not depend on past migrations. While there is evidence suggesting that, in reality, migration is a complex process driven by the cumulative effect of multiple factors through history, our assumptions help simplify the modelling process and make the interpretation of our models easier.

Essentially, the models are characterized by the transition probabilities of moving between cities, i.e.\ $m_{ij}$. The models can be more or less realistic depending on how these probabilities are defined. Sometimes, they can lead to unrealistic city size distributions in the long run, such as an equilibrium distribution where all the cities have the same population. However, most countries and regions display some hierarchy in their urban structure. Hence, a good model should be able to capture small and large cities simultaneously. 

We have explored a range of metrics that allow us to compare the distribution of city population sizes in an urban system. These metrics rely on the relative size of the cities but are independent of the number of cities and their absolute size. The size of a city alone does not say much about the role of a city in the urban hierarchy and, therefore, its degree of attractiveness for migrants \cite{SecodaryCitiesAfrica}. For example, China has more than 300 cities with more than one million inhabitants, so a city with one million people would be considered small. However, in Central Africa, a city with one million inhabitants is large enough to be a capital, such as N'Djamena, Libreville, Bangui, and Malabo in the case of Chad, Gabon, Central African Republic, and Equatorial Guinea, respectively \cite{SecodaryCitiesAfrica}. 

Reasons for moving to one city can be modelled with different techniques. Other chapters analyse ways to model the probability $\Pi$, depending on city size and distance between cities.

\section*{Acknowledgements}

This research is funded by the Federal Ministry of the Interior of Austria (2022-0.392.231) and by the UK Economic and Social Research Council (project number ES/Y010787/1).

\bibliographystyle{unsrt}

\end{document}